\documentclass[showpacs,amsmath,reprint,showkeys,pra,longbibliography]{revtex4-2}
\usepackage{dcolumn}    % Align table columns on decimal point
\usepackage{bm}         % bold math
\usepackage{ifpdf}
\usepackage{amssymb,lineno,amsfonts}
\usepackage{graphicx}   % Include figure files
\usepackage{bbm}
\usepackage{mathrsfs}
\usepackage{upgreek}
\usepackage{epstopdf}
\usepackage{setspace}
\usepackage{hyperref}
\usepackage[matrix,frame,arrow]{xypic}
\usepackage{float}
\usepackage{soul}
\usepackage{tikz}
\usepackage{natbib}
\usepackage{color} 
\usepackage{subfigure}
\usepackage{tikz}
\usepackage[]{qcircuit}

\definecolor{med-blue}{RGB}{25,25,112} 
\hypersetup{colorlinks, linkcolor={blue},citecolor={blue}, urlcolor={black}}
\newcommand{\ket}[1]{\vert{#1}\rangle}

\newcommand{\outpr}[2]{\vert{#1}\rangle\langle{#2}\vert}

\newcommand{\expv}[1]{\langle{#1}\rangle}

\newcommand{\proj}[1]{\outpr{#1}{#1}}

\newcommand{\dmel}[2]{\langle{#1}\vert{#2}\vert{#1}\rangle}

\newcommand{\tr}{\mathrm{Tr}}

\begin{document}
\title{NMR investigations of quantum battery using star-topology spin systems}
    \author{Jitendra Joshi}
    \email{jitendra.joshi@students.iiserpune.ac.in}
	\author{T S Mahesh}
	\email{mahesh.ts@iiserpune.ac.in}
	\affiliation{Department of Physics and NMR Research Center,\\
		Indian Institute of Science Education and Research, Pune 411008, India}

\begin{abstract}
%Recent proposals on using multi-level quantum systems as energy storage devices has led to an exciting possibility of realizing quantum batteries. 
Theoretical explorations have revealed that quantum batteries can exploit quantum correlations to achieve faster charging, thus promising exciting applications in future technologies.  Using NMR architecture, here we experimentally investigate various aspects of quantum battery with the help of nuclear spin-systems in star-topology configuration.  We first carry out numerical analysis to study how charging a quantum battery depends on the relative purity factors of charger and battery spins.  By experimentally characterizing the state of the battery spin undergoing charging, we  estimate the battery energy as well as the \textit{ergotropy}, the maximum amount of work that is unitarily available for extraction.  The experimental results thus obtained establish the quantum advantage in charging the quantum battery.   We propose using the quantum advantage, gained via quantum correlations among chargers and battery, as a measure for estimating the size of the correlated cluster. We develop a simple iterative method to realize asymptotic charging that avoids oscillatory behaviour of charging and discharging.  Finally, we introduce a load spin and realize a charger-battery-load circuit and experimentally demonstrate battery energy consumption after varying duration of battery storage, for up to two minutes.
\end{abstract}

\maketitle

\section{Introduction} 
Recent advances in quantum technologies are revolutionizing the world with novel devices such as quantum computers, quantum communication, quantum sensors, and a host of other quantum-enhanced applications \cite{nielsen2002quantum,lo1998introduction}. 
The latest additions include
quantum engines  \cite{quan2007quantum,goswami2013thermodynamics}, quantum diode \cite{palacios2018atomically,nakamura1996ingan}, quantum transistor \cite{geppert2000quantum}, as well as quantum battery, an energy-storing device  \cite{andolina2018charger,andolina2019extractable,aaberg2013truly} that is capable of exploiting quantum superpositions  \cite{alicki2013entanglement,
binder2015quantacell,campaioli2017enhancing,ferraro2018high,andolina2019quantum,campaioli2018quantum}.  
While quantum batteries open up novel applications, they are also exciting from the point of view of quantum thermodynamics \cite{2016,kosloff2013quantum,deffner2019quantum}, a rapidly emerging field that extends thermodynamical concepts to the quantum regime.
It has been theoretically established that quantum batteries can exhibit faster charging in a collective charging scheme that exploits quantum correlations \cite{kamin2020entanglement,binder2015quantacell,campaioli2017enhancing}. Recently quantum batteries with various models showing quantum advantages have been introduced \cite{monsel2020energetic,kamin2020non}.  They include  quantum cavity \cite{pirmoradian2019aging,ferraro2018high,zhang2018enhanced,andolina2019extractable,crescente2020charging,julia2020bounds,mohan2021reverse,niedenzu2018quantum,caravelli2020random}, spin chain
\cite{le2018spin,ghosh2020enhancement,PhysRevA.103.033715,rossini2019many,zakavati2021bounds,ghosh2021fast,santos2020stable}, Sachdev-Ye-Kitaev model \cite{rossini2020quantum,rosa2020ultra}, and quantum
oscillators \cite{andolina2019quantum,zhang2019powerful,andolina2018charger,chen2020charging}.
There also have been a few experimental investigations of quantum battery, such as the cavity assisted charging of an organic quantum battery  \cite{quach2020organic}.

Here we describe an experimental exploration 
of quantum batteries formed by nuclear spin-systems of different sizes in star-topology configuration.  Although, one can consider various other configurations, we find the star-topology systems to be particularly convenient for this purpose for the reasons mentioned in the review \cite{Mahesh_2021}.
Using NMR methods, we study various aspects of quantum battery by experimentally characterizing its state via quantum state tomography.  Thereby we monitor building up of battery energy during collective charging and establish the quantum speedup.  
We also estimate the quantity \textit{ergotropy}, that quantifies the maximum extractable work. By numerically quantifying quantum correlation in terms of entanglement entropy as well as discord, we reconfirm the involvement of correlations in yielding the quantum speedup. We therefore propose using the quantum speed to estimate size of the correlated cluster.  We find this method to be much simpler compared to spatial phase-encoding method \cite{pande2017strong} or the temporal phase-encoding method (eg. \cite{krojanski2006reduced}).
Unlike classical batteries,  charging of a quantum battery is oscillatory, i.e., the quantum battery starts discharging after reaching the maximum charge.  Recent theoretical proposals to realize a stable non-oscillatory charging were based on either adiabatic protocol \cite{santos2019stable} or shortcut to adiabaticity \cite{dou2022highly}.  Here we propose and demonstrate a simple iterative procedure to realize asymptotic charging based on the differential storage times of the charger and battery spins.  Finally, we describe implementing the Quantum Charger-Battery-Load (QCBL) circuit.  A similar circuit has recently been theoretically discussed in Ref. \cite{santos2021quantum}.  Using a 38-spin star-system we experimentally demonstrate QCBL circuit with battery  storage up to two minutes before discharging energy on to the load spin.

The article is organised as follows. In Sec. \ref{theory}, we describe the theoretical modeling of quantum battery and describe the numerical analysis of battery performance in terms of relative purity factors of charger and battery spins.  
In Sec. \ref{sec:expt}, we describe the following experimental studies on quantum battery.  The study of quantum advantage and ergotropy is reported in Sec. \ref{sec:qadv}.  The proposal to use quantum advantage as a measure of cluster size is discussed in Sec. \ref{sec:clustersize}.  The scheme to avoid oscillatory charging is described in Sec. \ref{sec:asymcharge}.  Finally, we describe implementation of the QCBL circuit in Sec. \ref{sec:qcbl} before summarizing in Sec. \ref{Conclusion}.

\begin{figure}
\centering
\hspace*{-0.1cm} (a) \hspace*{2.1cm} (b) \\
\centering
\includegraphics[trim=3cm 6cm 4cm 4cm,width=7.5cm,clip=]{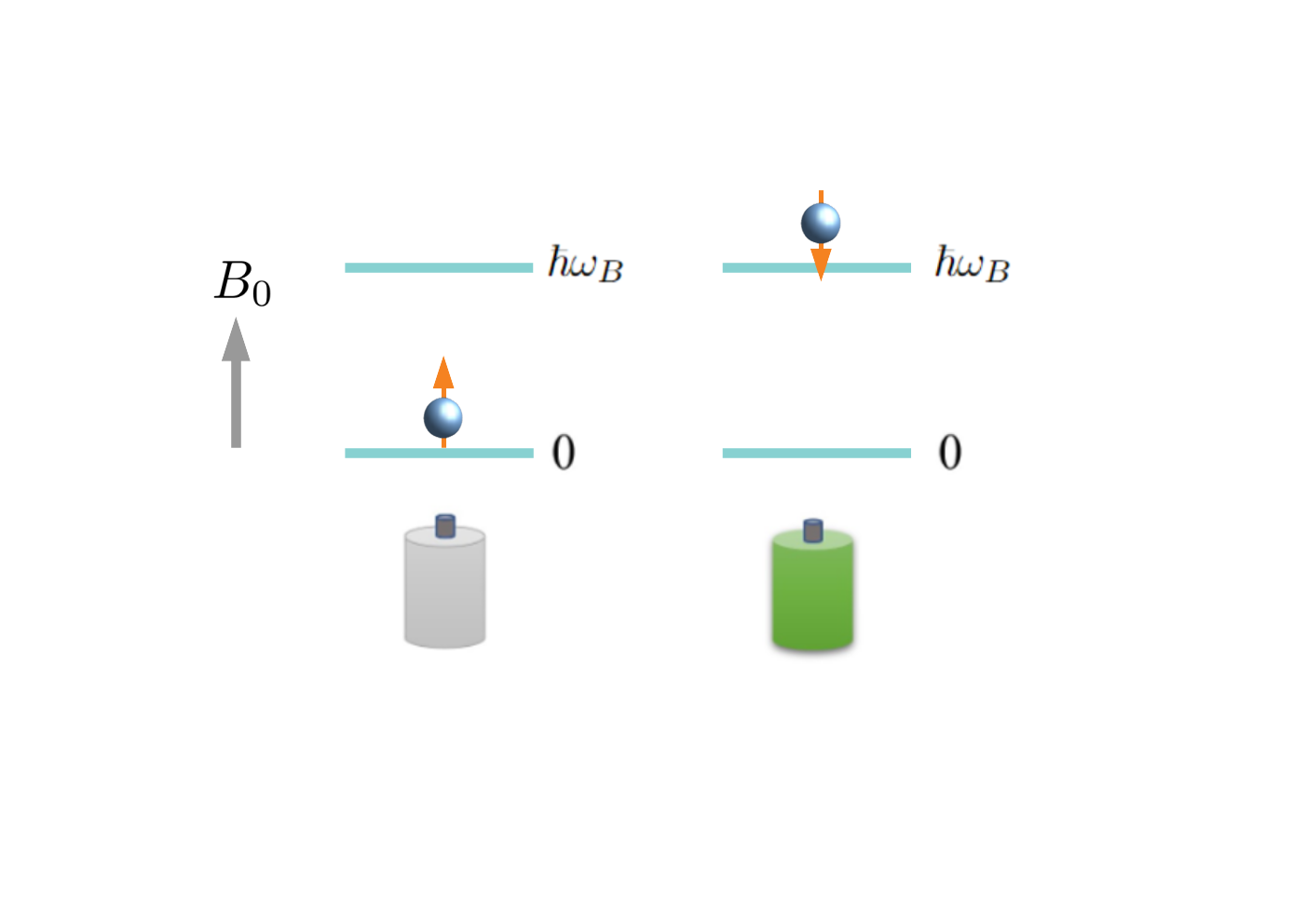}
\caption{\label{fig:qb}A single spin-1/2 particle in an external magnetic field $B_0$ as a quantum battery. The ground state (a) and excited state (b) correspond respectively to uncharged and charged states of the battery. }
\end{figure}

\section{Theory}
\label{theory}
\subsection{A nuclear spin-battery} 
The simplest quantum battery (B) consists of a two-level quantum system, like a spin-1/2 particle placed in a magnetic field (Fig. \ref{fig:qb}).
Here, the ground state $\ket{0}$ is modeled as a discharged or empty battery, while the excited state $\ket{1}$ is modeled as the fully charged battery.
The spin battery can be charged either directly using an external drive \cite{binder2015quantacell,ferraro2018high} or indirectly via an ancillary spin, called charger spin (C) \cite{le2018spin,santos2021quantum}. 
Let us now consider the B-C spin system.
Each of the two spins are governed by their local Hamiltonians $H_B$ and $H_C$, respectively, which for the sake of simplicity, are chosen to have zero ground-state energy.
Moreover, we assume that the quantum system at an initial time $t = 0$ is in a factorized state 
\begin{align}
\rho_{BC}(0) = \proj{0}_B\otimes\proj{1}_C,   
\label{eq:rhobc1chargerpure}
\end{align}
with $\proj{1}_C$  being the excited state of the charger.

We now introduce a coupling Hamiltonian $H_{BC}(t)$ between B and C, in order to transfer as much energy as possible from the charger to the battery over a finite charging duration $\tau$. Under the global Hamiltonian of the system BC
\begin{align}
\centering H(t) = H_B + H_C+ H_{BC}(t),
\end{align}
the joint system evolves as
\begin{align}
\label{unitary}
\rho_{BC}(\tau) &= U(\tau) \rho_{BC}(0)U^{\dagger}(\tau)
\nonumber \\
&~\mbox{with}~ U(\tau) = Te^{-i\int_0^\tau dt H(t)},
\end{align} 
%with $U(\tau) = Te^{-i\int_0^\tau dt H(t)}$,
where $T$ is the time-ordering operator.  The instantaneous state of battery $\rho_B(\tau) = \tr_C(\rho_{BC}(\tau))$ is obtained by tracing out the charger.
The goal is to maximize the local energy of the battery 
\begin{align}
E_B^\mathrm{max} = E_B(\overline{\tau}) = \tr(\rho_{B}(\overline{\tau}) H_B), 
\end{align}
with the shortest possible charging time $\overline{\tau}$. 
For a given maximum energy charged $E_B^\mathrm{max}$, the charging power is defined as 
\begin{align}
P = E_B^\mathrm{max}/\overline{\tau}.
\end{align}

\begin{figure}
\centering
\includegraphics[trim=0cm 0cm 0cm 0cm,width=7.5cm,clip=]{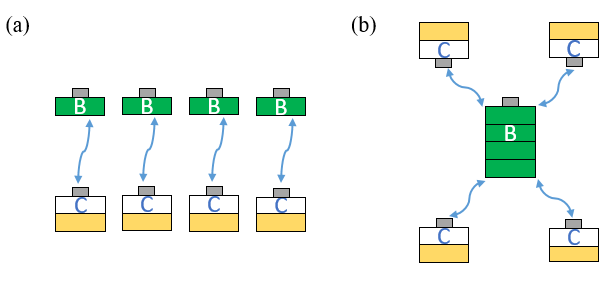}
\caption{\label{fig:battery}Two charging schemes: (a) parallel charging scheme where a single battery  is charged by an individual charger and (b) the collective charging scheme where a single battery is charged by multiple chargers. }
\end{figure}

We now discuss two charging schemes, parallel and collective \cite{binder2015quantacell,ferraro2018high,le2018spin} as illustrated in Fig. \ref{fig:battery}.  In  parallel charging scheme (Fig. \ref{fig:battery}(a)), each of the $N$ batteries is independently charged to a maximum energy $E_B^\mathrm{max}/N$ by one of the $N$ chargers over a duration $\overline{\tau}_1$. Conversely, in the collective charging scheme (Fig. \ref{fig:battery}(b)), all the batteries together form a battery-pack that is charged to a maximum energy $E_B^\mathrm{max}$ simultaneously by $N$ chargers over a duration $\overline{\tau}_N$.  The latter scheme exploits quantum correlations and hence is more efficient \cite{ferraro2018high,binder2015quantacell}.
Let $P_1$ and $P_N$ be the the charging powers of the parallel and collective charging schemes respectively. The quantum advantage of collective charging is defined as  \cite{campaioli2017enhancing}
\begin{eqnarray}
\label{gamma}
\Gamma \equiv \frac{P_N}{P_1}
= \frac{E_B^\mathrm{max}/\overline{\tau}_N}{N(E_B^\mathrm{max}/N)/\overline{\tau}_1} = \frac{\overline{\tau}_1}{\overline{\tau}_N}.
\end{eqnarray}

We may also characterize the state of the battery during  charging in terms of ergotropy, or the maximum work that can be extracted \cite{allahverdyan2004maximal}.
Following Refs. \cite{allahverdyan2004maximal,pusz1978passive,campaioli2018quantum}, the ergotropy of a battery at time $\tau$ is given by
\begin{eqnarray}
\label{eq:ergo}
{\cal E}(\rho_{B}(\tau)) = E_B(\rho_{B}(\tau)) - E_B(\rho_{B}^p(\tau)),
\end{eqnarray}
 where $E_B(\rho) = \tr(\rho H_B)$ is the  energy of the state $\rho$ and $\rho_{B}^{p}(\tau)$ is the passive state corresponding to  $\rho_{B}(\tau)$. A passive state, or a zero-ergotropy state, is the one from which no work can be extracted by using unitary methods \cite{allahverdyan2004maximal,pusz1978passive}.  To construct the passive state, we first spectrally decompose the state $\rho_{B}(\tau)$ and Hamiltonian $H_B$ as
 \begin{align}
\rho_{B}(\tau) &= \sum_{j} r_j \proj{r_j},    ~\mbox{where}~ r_1 \ge r_2 \ge \cdots, \mbox{and}
\nonumber \\
H_B &= \sum_k E_k \proj{E_k} ~\mbox{where}~ E_1 \le E_2 \le \cdots.
\label{eq:rhobhb}
 \end{align}
The passive state is diagonal in the energy basis formed by pairing descending order of populations $r_j$ with ascending order of energy $E_j$ levels, i.e.,
\begin{align}
\rho_{B}^p(\tau) = \sum_j r_j \proj{E_j}.
\label{eq:permutedrhob}
\end{align}
Note that the energy of the passive state is
\begin{align}
E_B(\rho_{B}^p(\tau)) = \sum_j r_j E_j.
\end{align}
For a single spin battery described in Fig. \ref{fig:qb}, the eigenvalues of instantaneous state are of the form $(1\pm \epsilon)/2$ where $|\epsilon| \le 1$.  Therefore, 
\begin{align}
\rho_{B}(\tau) &= \frac{1+\epsilon}{2} \proj{0} + \frac{1-\epsilon}{2} \proj{1} ~~\mbox{and}
\nonumber \\
E_B(\rho_{B}(\tau)) &= \hbar\omega_B \frac{1-\epsilon}{2}.
\label{eq:rhobeb}
\end{align}
As long as $\epsilon \ge 0$, the ground state is still more populated than the excited state, and the battery remains in the passive state and ergotropy ${\cal E}(\rho_{B}(\tau)) = 0$.  After sufficient charging, $\epsilon$ becomes negative, and the passive state changes to
\begin{align}
\rho_{B}^p(\tau) &= \frac{1-\epsilon}{2} \proj{0} + \frac{1+\epsilon}{2} \proj{1}, 
\nonumber \\
E_B(\rho_{B}^p(\tau)) &= \hbar\omega_B \frac{1+\epsilon}{2},
\nonumber \\
\mbox{and ergotropy}~ {\cal E}(\rho_{B}(\tau)) &= -\epsilon \hbar\omega_B ~~(\epsilon \le 0).
\end{align}
For $|\epsilon| \ll 1$ we find that the dimensionless ratio
\begin{align}
\frac{{\cal E}(\rho_{B}(\tau))}{-\epsilon E_B(\rho_{B}(\tau))} &= \frac{2}{1-\epsilon} \approx 2. 
\label{eq:ergofactor}
\end{align}
In the following we describe the topology of the spin-systems used in our experiments.

\subsection{Star-topology network} 
\label{Star-topology network}
\begin{figure}[h]
\centering
\includegraphics[trim={0cm 0.7cm 0cm 0cm},clip,width=7cm]{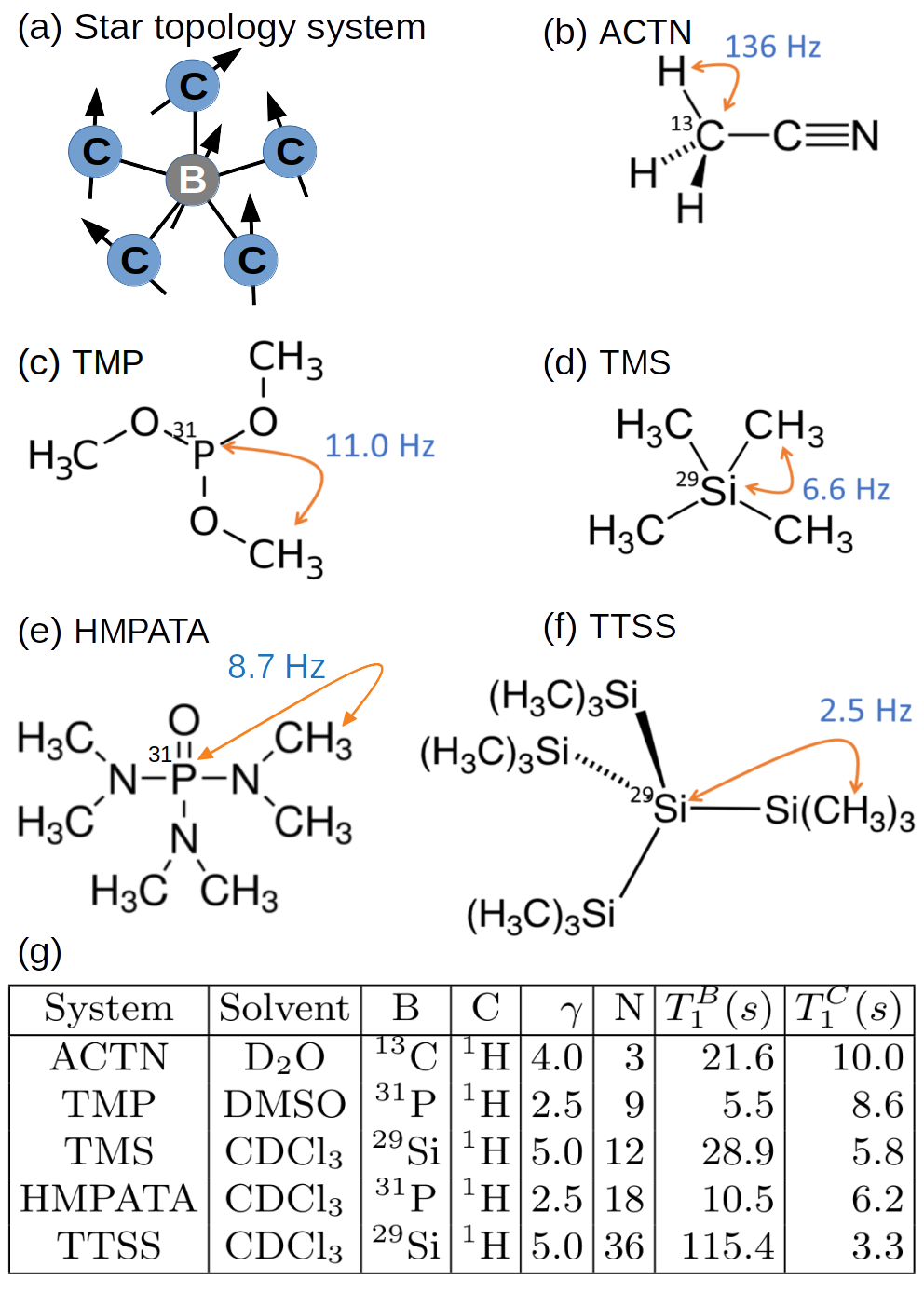}
\caption{\label{fig:NMR}(a) Star-topology configuration showing the central battery spin symmetrically surrounded by charger spins.  (b-f) The star-topology nuclear spin-systems studied in this work. The strength $J$ of battery-charger interaction for each system is shown with the molecular structure, while other details are tabulated in (g). Note that all the nuclei considered here (B and C) are spin 1/2 nuclei.}
\end{figure}

We now consider the star-topology network in which a single central battery-spin uniformly interacts with a set of $N$ indistinguishable charger spins \cite{Mahesh_2021} as illustrated in Fig. \ref{fig:NMR} (a).  Quantum battery in this configuration has been studied theoretically very recently \cite{PhysRevB.104.245418}. The spin-systems with $N = 3, 9, 12, 18,~\&~ 36$ studied in this work are shown in Fig. \ref{fig:NMR} (b-f). 

We consider the local Hamiltonians for the battery and charger to be
\begin{align}
H_B = \hbar \omega_B(1/2-S_z)
~\mbox{and}~
H_C = \hbar \omega_CI_z. \end{align}
Here $S_{x,y,z}$ represent the $x,y,z$-spin operators for the battery spin with Larmor frequency $\omega_B$, $I_{x,y,z} = \sum_{i=1}^N I_{x,y,z}^i$ represent the collective $x,y,z$-spin operators for the chargers with Larmor frequency $\omega_C = \gamma \omega_B$, where $\gamma$ is the relative gyromagnetic ratio.
Following Ref. \cite{andolina2019quantum}, we choose the interaction Hamiltonian,
\begin{align}
H_{BC}(t) &= \hbar 2\pi J  \left( S_x I_x + S_y I_y \right),
\end{align}
where $J \ll |\omega_{C(B)}|$ is the coupling constant between the battery and the charger spins. 

The spin-system is prepared in the thermal equilibrium state, which is in a generalized form of Eq. 
\ref{eq:rhobc1chargerpure}, i.e.,
\begin{align}
\rho_{BC}(0) &= \rho_B(0) \otimes \rho_C(0), ~\mbox{with}
\nonumber \\
\rho_B(0) &= \frac{1+\epsilon}{2} \proj{0}
+\frac{1-\epsilon}{2} \proj{1}
~\mbox{and}
\nonumber \\
\rho_C(0) &= \left(\frac{1-\gamma\epsilon}{2} \proj{0}
+\frac{1+\gamma\epsilon}{2} \proj{1}
\right)^{\otimes N},
\label{eq:rhobcrmixed}
\end{align}
where $\epsilon$ and $\gamma \epsilon$ are the purity factors of the battery and charger spins respectively.  Under the high-temperature approximation relevant for NMR conditions, $\epsilon \approx 10^{-5}$.

\begin{figure}[h]
\centering
\includegraphics[trim=0.6cm 0.5cm 0.6cm 0.5cm,width=8cm,clip=]{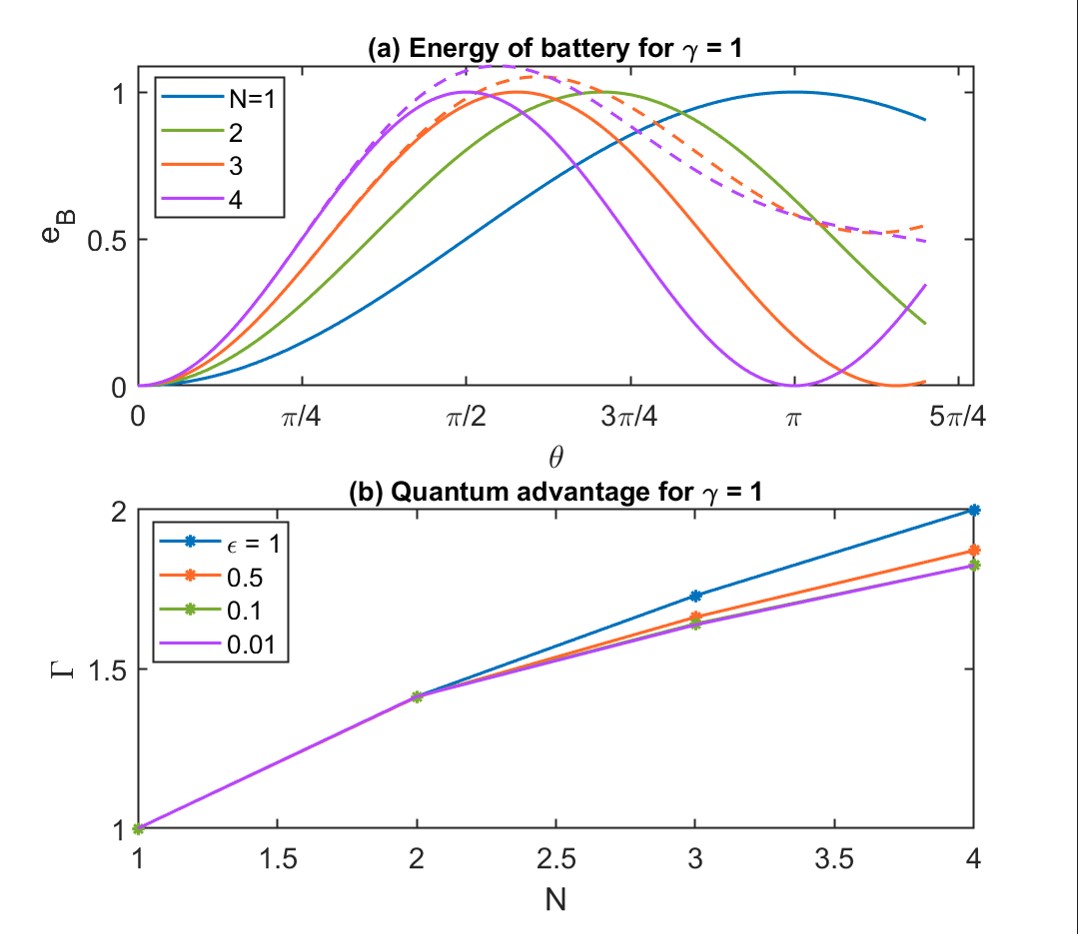}
\caption{\label{fig:QAdv} (a) Battery energy ${\tt e}_B$ versus charging phase $\theta = 2\pi J \tau$ for different number $N$ of chargers in pure (solid lines) as well as mixed (dashed lines; $\epsilon = 10^{-5}$) state cases.  (b) Quantum advantage $\Gamma$ versus $N$ for different  purity values $\epsilon$. }
\end{figure}

We evolve the whole system for a duration $\tau$ under the total Hamiltonian in the interaction frame defined by $U_\mathrm{IF}(t) = e^{-i(H_B+H_C) t /\hbar}$.
The dimensionless energy of the battery
\begin{align}
{\tt e}_B(\tau) = E_B(\tau)/\hbar \omega_B = \dmel{1}{\rho_B(\tau)}
\end{align}
is related to the normalized polarization of the battery
\begin{align}
m_B(\tau) = \expv{\sigma_z}_{\rho_B(\tau)}/\epsilon   
~~\mbox{via}~~
{\tt e}_B(\tau) = \frac{1-m_B(\tau)}{2}.
\label{mb2eb}
\end{align}
For the special case of pure state, i.e., $\epsilon = 1$ and also setting $\gamma=1$, we obtain the state and dimensionless energy as
\begin{align}
\label{eq:rho}
\rho_B(\tau) = \cos^{2}(\sqrt{N}\theta/2) \proj{0} + \sin^{2}(\sqrt{N}\theta/2) \proj{1} 
 \end{align}
\begin{align}
 {\tt e}_B(\tau) = \sin^{2}(\sqrt{N}\theta/2)~
 \mbox{in terms of}~\theta = 2\pi J \tau.
 \label{eq:ebpure}
 \end{align}
  The energy is maximized for $\overline{\theta} = \pi/\sqrt{N}$ at optimal time 
\begin{align}
\overline{\tau}_N &=
\frac{\overline{\theta}}{2\pi J} = 
\frac{1}{2J\sqrt{N}},~~
 \therefore ~~
\Gamma = 
\frac{\overline{\tau}_1}{\overline{\tau}_N} =
\sqrt{N},
\end{align} 
clearly predicting the quantum speed-up.
 The battery energy evolution for various numbers of charger spins are shown in Fig. \ref{fig:QAdv} (a). 
 Note that mixed state curves deviate from the pure state curves for $N \ge 3$.  Here, while ${\tt e}_B$ exceeds the pure state value of unity, the maximum charging takes longer duration.  The quantum advantage $\Gamma$ versus number of charger spins for $\gamma=1$ and various values of $\epsilon$ are shown in Fig. \ref{fig:QAdv} (b). 
In the following we discuss the experimental investigation of quantum battery.

\section{Experiments} 
\label{sec:expt}
\subsection{Establishing quantum advantage \label{sec:qadv}}
\begin{figure}
\centering
(a) \\
\centering
\includegraphics[trim={4cm 5.5cm 5cm 4cm},clip,width=7.5cm]{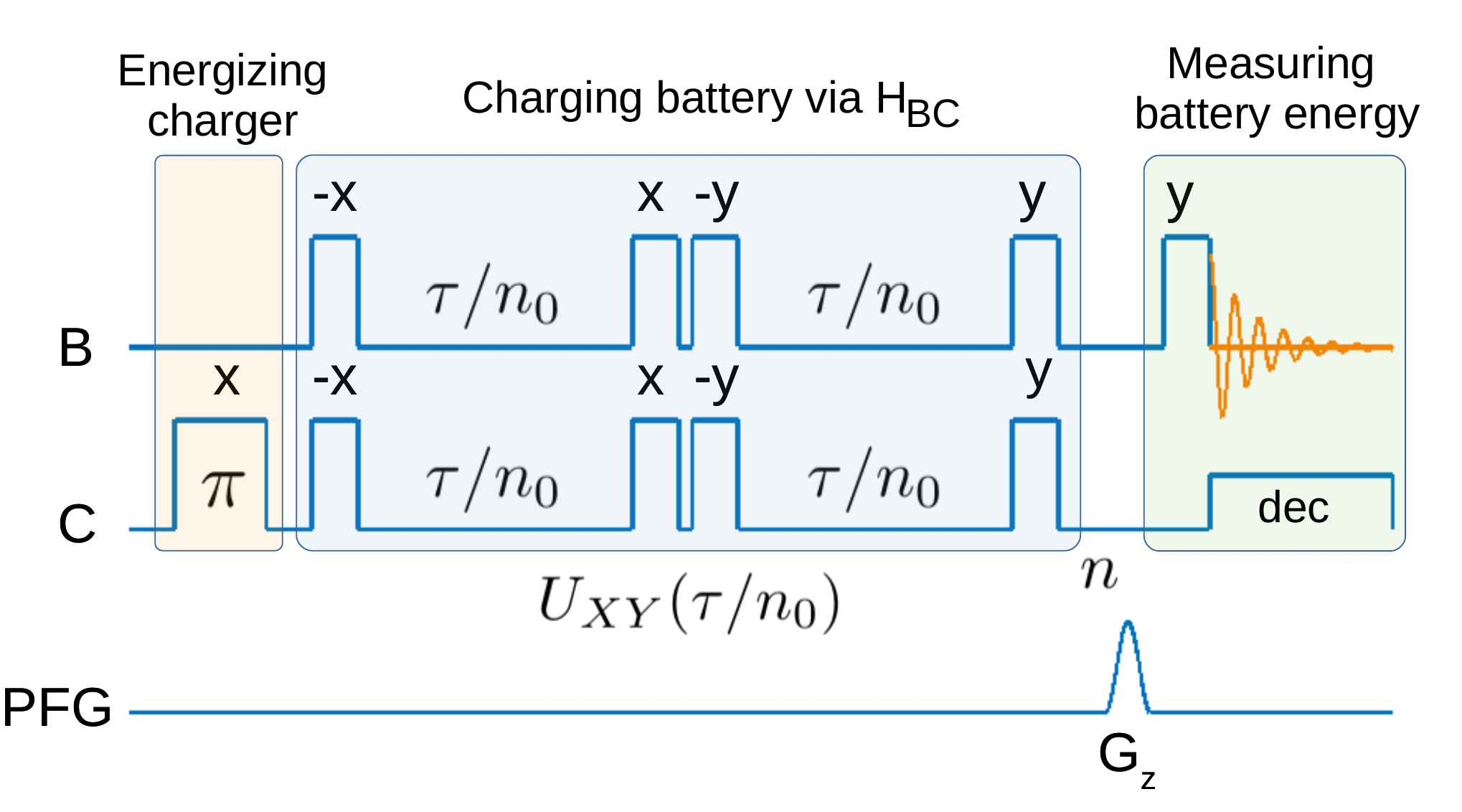} \\
\hspace*{0.3cm} (b) \hspace*{2.6cm} (c) \\ \includegraphics[trim={1.1cm 0cm 1.5cm 0.5cm},clip,width=7.5cm]{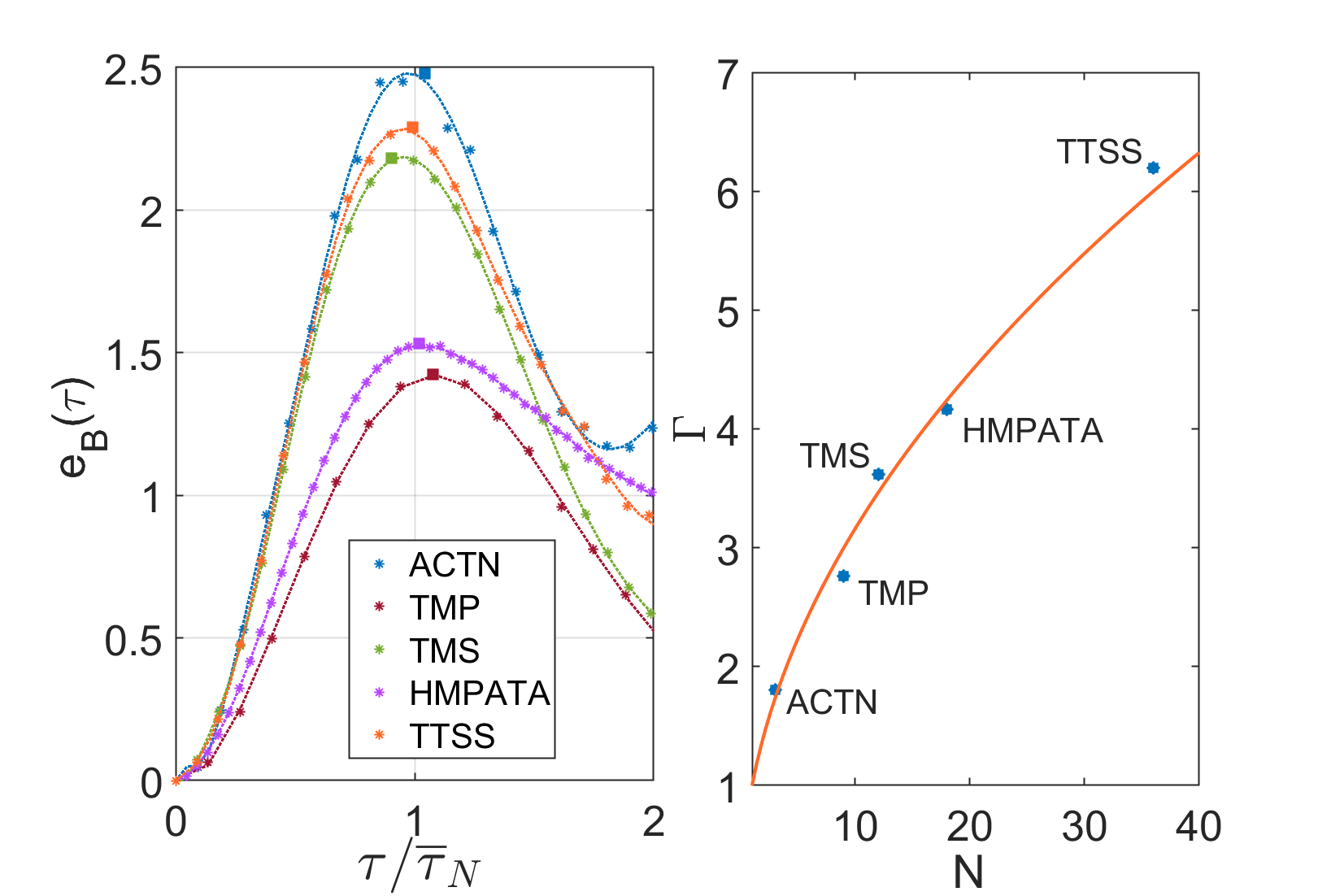}
\caption{\label{fig:rootN}
(a) The NMR pulse sequence for charging  quantum battery and measuring its energy.
The wide and narrow rectangular pulses correspond to $\pi$ and $\pi/2$ pulses respectively.  The shaped pulse in the lowest row corresponds to the pulsed-field-gradient (PFG) which dephases the coherences and retains populations.
(b) The dots correspond to experimentally measured battery energy values ${\tt e}_B$ versus normalized charging duration $\tau/\overline{\tau}_N$ for the five spin-systems shown in Fig. \ref{fig:NMR}. Here the solid lines are spline-fits to guide the eye.(c) Quantum advantage $\Gamma$ versus the number $N$ of charger spins showing $\sqrt{N}$ dependence.}
\end{figure}

Our first aim is to establish the quantum advantage described in section \ref{Star-topology network} using various systems shown in Fig. \ref{fig:NMR}.  The table containing information about the solvent, the relative gyromagnetic ratio ($\gamma$), and the T$_1$ relaxation time constant for each of the spin systems is shown in Fig. \ref{fig:NMR} (g).
All the experiments were carried out in a 500 MHz Bruker NMR spectrometer at an ambient temperature of 298 K. 
The NMR pulse-sequence for the experiments is shown in Fig. \ref{fig:rootN} (a).
Starting from thermal equilibrium state, we energize the charger spins by inverting their populations with the help of a $\pi$ pulse.  This is followed by the charging propagator
 \begin{align}
    U_{XY}(\tau/n_0)
    &= e^{-i H_{BC}\tau/n_0} 
    \nonumber \\
    &\approx Y\cdot ZZ \cdot Y^\dagger \cdot X\cdot ZZ \cdot X^\dagger.
\end{align}
Here, $X(Y) = e^{-i(S_{x(y)}+I_{x(y)})\pi/2}$ and
$ZZ = e^{-i S_zI_z \theta/m}
$.  Note that for $N \ge 2$, $[S_xI_x,S_yI_y] \neq 0$, and therefore we implement the interaction propagator via integral iterations  $n \in [0,n_0]$ of $U_{XY}(\tau/n_0)$ with sufficiently large $n_0$ such that $\tau/n_0 \ll 1/(2J)$. Finally, after dephasing spurious coherences with the help of a pulsed-field-gradient (PFG), we apply a  $\pi/2$ detection pulse and measure the battery polarization $m_B(\tau)$.  During the detection period, we decouple  charger spins using WALTZ-16 composite pulse sequence \cite{cavanagh1996protein}. 
\begin{figure}
\centering
\includegraphics[trim={0cm 0cm 0cm 0cm},clip,width=9cm]{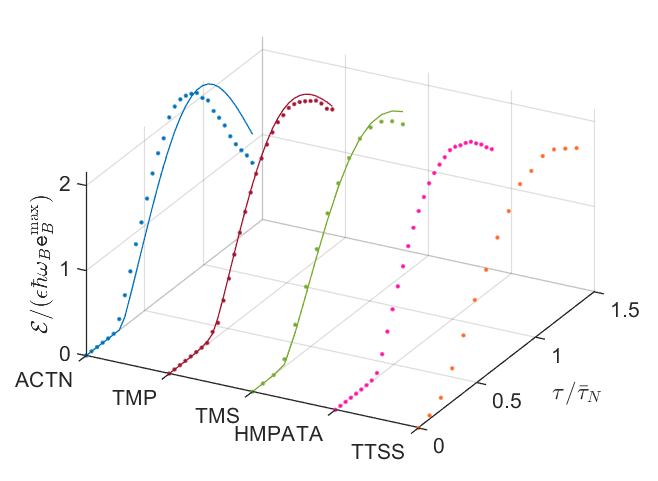}
\caption{\label{fig:ergotropy}The dots represent the experimentally estimated ergotropy of the battery-spin versus normalized charging duration $\tau/\bar{\tau}_N$ for all five spin-systems.  Here the ergotropy is scaled by $\epsilon \hbar \omega_B {\tt e}_B^{\mathrm{max}}$ (see Eq. \ref{eq:ergofactor}), where ${\tt e}_B^{\mathrm{max}}$ is taken from Fig. \ref{fig:rootN}(b) highlighted by bigger rectangular dots.  The solid lines in small spin-systems represent the theoretical fits accounting also for the experimental nonidealities.  
}
\end{figure}

The experimentally measured battery energy ${\tt e}_B$ estimated from $m_B$ using Eq. \ref{mb2eb} for all five spin-systems shown in Fig. \ref{fig:NMR} are plotted versus normalized charging duration $\tau/\overline{\tau}_N$ in Fig. \ref{fig:rootN} (b).  For an ideal pure-state system, we expect the maximum energy storage at $\tau/\overline{\tau}_N = 1$.  On the other hand, for mixed state systems with $N \ge 3$, $\tau/\overline{\tau}_N$ slightly overshoots the unit value.  However, in practical systems, the charging dynamics is affected by the experimental imperfections such as RF inhomogeneity (RFI), off-set and calibration errors, etc.  In spite of these issues, the results shown in Fig. \ref{fig:rootN} (b) for all the systems show a remarkable agreement with the expected maximum charging duration at $\overline{\tau}_N$.  The corresponding quantum advantage $\Gamma$ for all the systems are plotted versus the number $N$ of charger spins in Fig. \ref{fig:rootN} (c), where the solid line corresponds to the theoretically expected $\sqrt{N}$ function.  Clearly, we observe a significant quantum advantage ranging from about 1.5 to over 6.

We now explain the experimental measurement of ergotropy for the subsystem consisting only the battery spin.  To this end, we carry out the complete quantum state tomography \cite{nielsen2002quantum} of the battery spin while tracing out the charger spins using heteronuclear composite pulse decoupling.  After reconstructing the density matrix $\rho_B(\tau)$ we use Eqs. (\ref{eq:ergo}-\ref{eq:permutedrhob}) to estimate the ergotropy value. The dots in Fig. \ref{fig:ergotropy} represent the experimentally estimated ratio of ergotropy to maximum energy (see Eq. \ref{eq:ergofactor}) plotted versus the normalized charging time $\tau/\bar{\tau}_{N}$.  Here the solid lines are theoretical fits accounting for experimental nonidealities such as RFI, relaxation effects, etc.
As explained after Eq. \ref{eq:rhobeb}, the battery spin remains in a passive state and exhibits zero ergotropy until its populations are saturated.  Ideally for $\gamma=1$, the saturation occurs at time $1/(4J \sqrt{N})$ (follows from Eq. \ref{eq:rho}), while for $\gamma \ge 1$, it occurs earlier.  
Once the battery-spin populations begin to invert, the ergotropy ratio starts building up towards the value 2 (see Eq. \ref{eq:ergofactor}) and reaches its maximum at normalized charging time $\tau/\bar{\tau}_N = 1$.  Thus, once again we observe the quantum advantage in charging of quantum battery.

\subsection{Determining size of the correlated cluster \label{sec:clustersize}}
\begin{figure}[h]
\centering
\includegraphics[trim={1.5cm 0cm 1.5cm 0.5cm},clip,width=6.5cm]{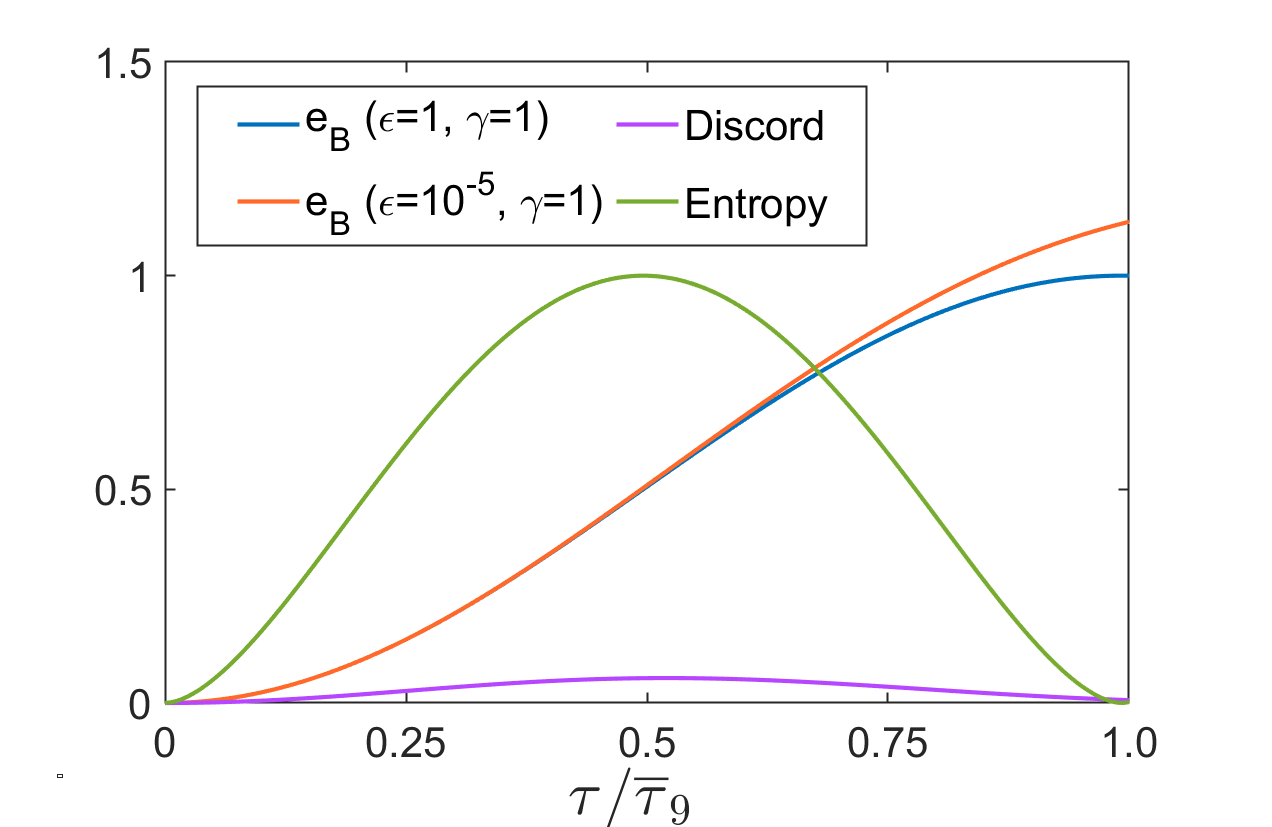}
\caption{\label{fig:entropy}
Numerically calculated battery energy (with pure and mixed states),
entanglement entropy (for pure state; $\epsilon = 1$, $\gamma = 1$), and quantum discord (for mixed state; $\epsilon = 10^{-5}$, $\gamma = 1$) versus the normalized charging duration $\tau/\overline{\tau}_9$ for $N=9$ star-system involving a single battery spin and nine charger spins.  
}
\end{figure}

It has been shown that quantum correlation plays a key role while charging quantum battery via collective mode \cite{binder2015quantacell}.  The same holds true for charging in the star-topology system.  In Fig. \ref{fig:entropy}, we plot  entanglement entropy as well as quantum discord against the normalized charging time $\tau/\overline{\tau}_9$ for a star-system with $N=9$ charger spins.  For reference we also show the charging energy ${\tt e}_B$ for both pure (with $\epsilon = 1$, $\gamma = 1$) and mixed state (with $\epsilon = 10^{-5}$, $\gamma = 1$).  To evaluate entanglement entropy we traced out charger spins, and evaluated the von Neumann entropy of the battery state.  For evaluating quantum discord, we used the two-spin reduced state obtained by tracing out all spins except the battery spin and one of charger spins.
We find that the maximum correlation is reached at $\tau/\overline{\tau}_9 = 0.5$, i.e., at half the maximum charging period.  Both entanglement entropy and discord vanish at maximum charging period, i.e., $\tau/\overline{\tau}_9 = 1$, and the spins get uncorrelated  \cite{binder2015quantacell,alicki2013entanglement}.  Since (i) the quantum advantage is linked to the generation of correlated state \cite{binder2015quantacell} and (ii) the maximum charging period depends on the size of the correlated cluster,  here we propose to use $\Gamma^2+1$ as an estimate for size of the correlated cluster.  This is justified by the good agreement between the theory and experiment for all the five systems investigated in Fig. \ref{fig:rootN} (b) and (c).
For example, the experimentally obtained value $\Gamma \approx 6$ for TTSS matches with the correlated cluster of 37 spins.

\subsection{Asymptotic charging
\label{sec:asymcharge}}
\begin{figure}[h]
\hspace*{0.8cm}(a)\\
\centering
\includegraphics[trim={4cm 6cm 6cm 5cm},clip,width=7cm]{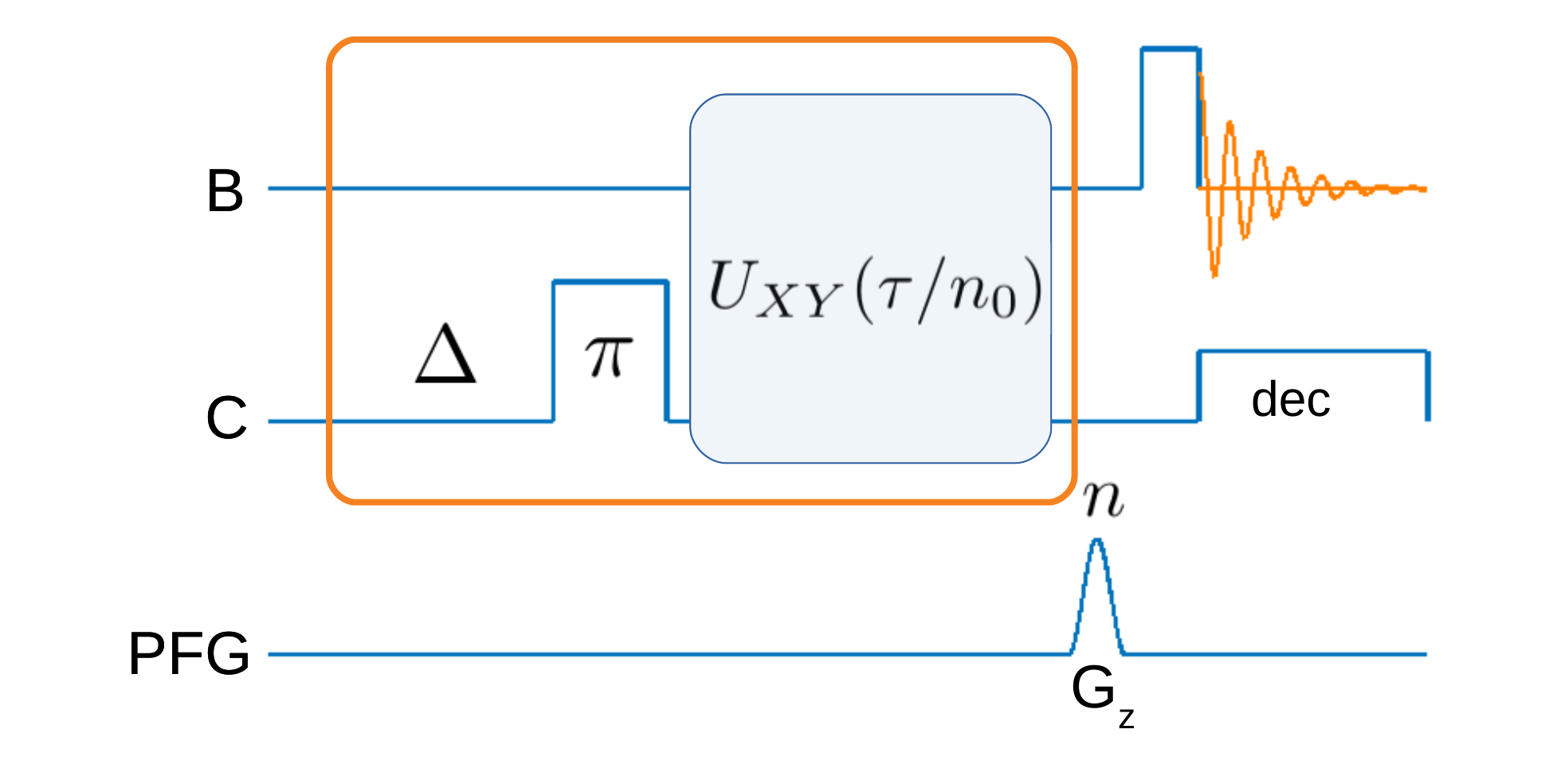} \\
\hspace*{0.8cm} (b) \hspace*{3.4cm} (c) \includegraphics[trim={2cm 0.5cm 1cm 0.5cm},clip,width=8cm]{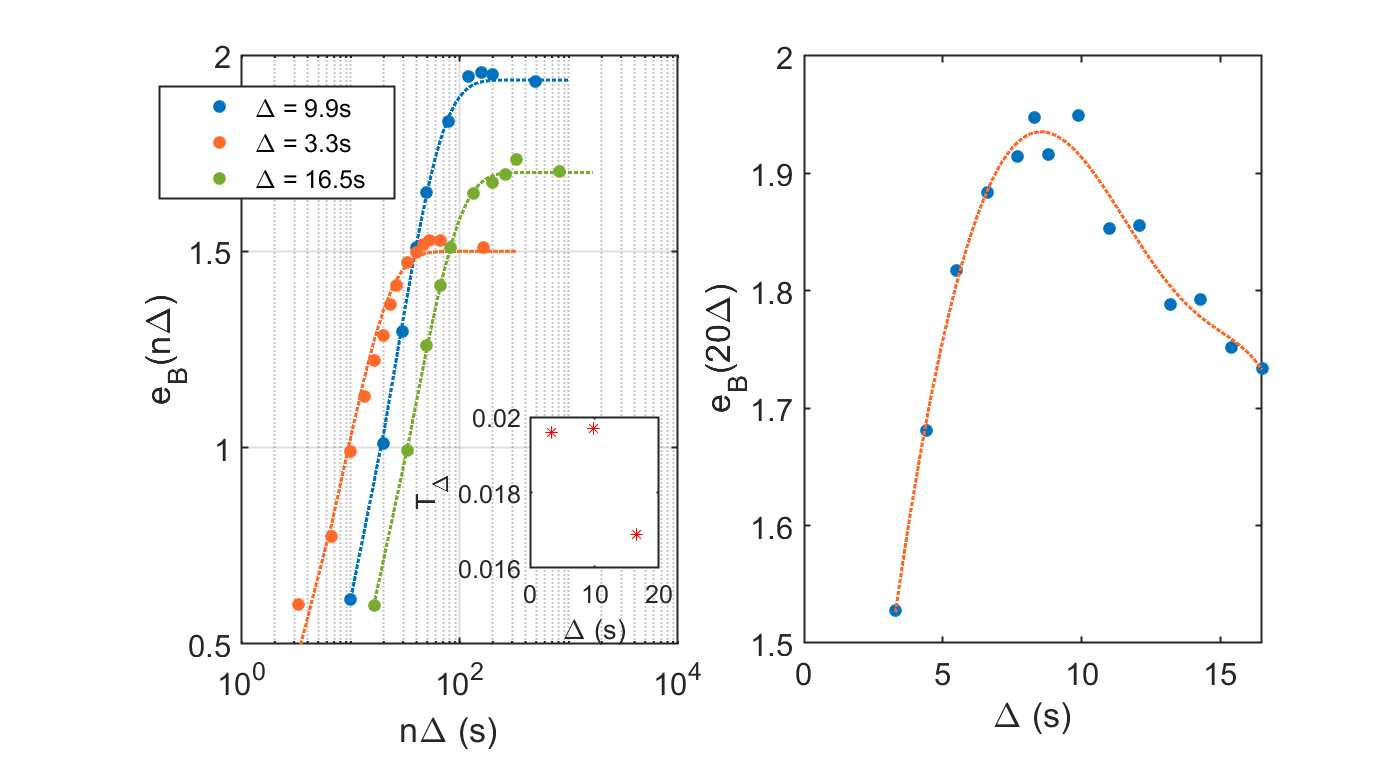}
\caption{\label{fig:Asym_charg}
(a) The NMR pulse sequence for asymptotic charging of a quantum battery.  (b) Battery energy ${\tt e}_B$ versus charging duration $n\Delta$ for three values of delay $\Delta$.   Here the dashed lines represent the fits to asymptotic charging functions as described in the text. The charging time-constants for these three cases are plotted in the inset. (c) Battery energy at saturation ${\tt e}_B(20 \Delta)$ (after $n=20$ iterations) versus the delay $\Delta$ showing the optimal delay range from 7.5 s to 10 s.  Here the dashed line is a spline curve fit to guide the eye.}
\end{figure}

We now propose a simple method to avoid oscillatory charging and realize an asymptotic charging that keeps the quantum battery from discharging.
The method relies on the differential storage times of the charger and the battery spins, i.e., $T_1^B \gg  T_1^C$.  It involves iteratively re-energizing the chargers followed by transferring the charge to the quantum battery after a carefully chosen delay.  The scheme for the asymptotic charging is described by the pulse-sequence shown in Fig. \ref{fig:Asym_charg} (a).  It involves a delay $\Delta$ before energizing the charger followed by charging the battery. However, unlike the unitary scheme described in section \ref{sec:qadv}, here the entire process including waiting time, re-energizing of the battery, and charging is iterated.  The experimentally measured battery energy ${\tt e}_B$ of the asymptotic charging with TTSS system are shown by dots in Fig. \ref{fig:Asym_charg} (b), wherein the dashed lines represent the fits to asymptotic charging functions ${\tt e}_B(n\Delta) = {\tt e}_B^\Delta(1-e^{-n\Delta/T_\Delta})$. 
Note that for TTSS, $T_1^B = 115.4$ s which is much longer than $T_1^C = 3.3$ s (see Fig. \ref{fig:NMR} (g)).
The estimated values of the charging time-constants $T_\Delta$ is plotted versus $\Delta$ in the inset of Fig. \ref{fig:Asym_charg} (b).  It is clear that there is an optimal delay time $\Delta$ for which we observe maximum charging.  
Therefore, we monitored the saturation charging, i.e., ${\tt e}_B(20\Delta)$ versus the delay time $\Delta$ as shown in Fig. \ref{fig:Asym_charg} (c).  For TTSS, we find the optimal delay ranges from 7.5 s to 10 s, to asymptotically achieve  over 85 \% charging compared to  the simple unitary method described in  section. \ref{sec:qadv}.

\begin{figure}[h]
\centering
(a)\\
\centering
\includegraphics[trim={0cm 8.5cm 0cm 1cm},clip,width=8cm]{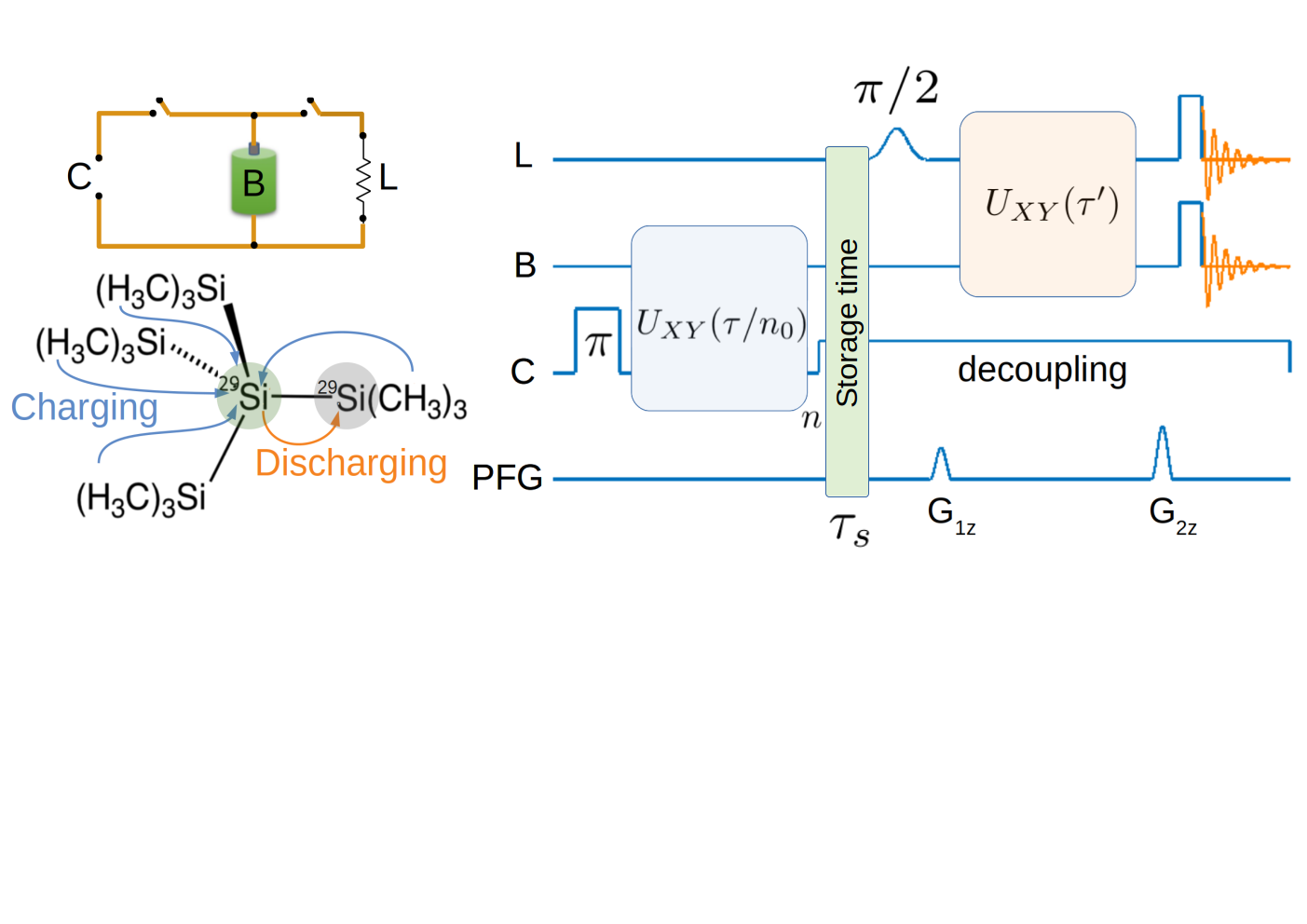}
\\
\hspace*{0.8cm} (b) \hspace*{3.4cm} (c) 
\includegraphics[trim={1cm 0.3cm 1cm 0.5cm},clip,width=8.4cm]{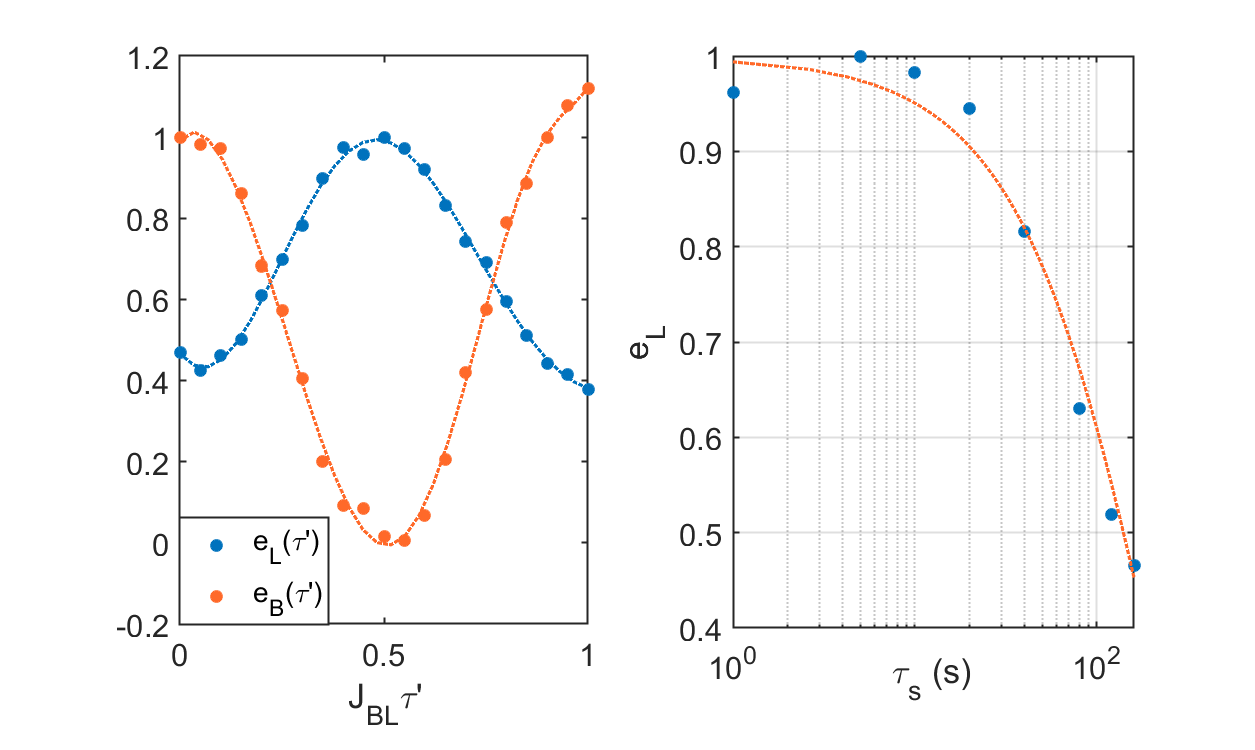}
\caption{\label{fig:CD1}(a) The QCBL circuit and its implementation in the 38-spin star-topology system (left) and the NMR pulse sequence for the QCBL circuit (right).  Here the dashed lines are spline curve fits to guide the eye. (b) The energy of battery (${\tt e}_B$) and load (${\tt e}_L$) versus discharging parameter $J_{BL}\tau'$.  (c) The energy of the load (${\tt e}_L$) extracted from the battery after a storage time $\tau_s$.  The dashed line is an exponential fit as discussed in the text.
}
\end{figure}

\subsection{Quantum Charger-Battery-Load (QCBL) Circuit \label{sec:qcbl}}
Now we describe the QCBL circuit consisting of charger (C), battery (B), as well as a load (L).   Here we again use TTSS system, and consider  all the proton spins together as charger,
the central $^{29}$Si spin as the battery, and the peripheral $^{29}$Si spin as the load.   Given the $5\%$ natural abundance of $^{29}$Si, the probability of both central and one of the four peripheral silicon nuclei to be $^{29}$Si isotope is $0.2\%$.
In this system, the strength of the $^{29}$Si-$^{29}$Si interaction, i.e., $J_{BL} = 52.4$ Hz. 
The  QCBL circuit and the corresponding spin labeling are illustrated on the left of Fig. \ref{fig:CD1} (a). The NMR pulse-sequence for QCBL is shown on the right side of Fig. \ref{fig:CD1} (a).  We first charge the battery (B) as described in Sec. \ref{sec:qadv} and switch-off the C-B interactions by decoupling the charger spins throughout.
Subsequently, we can introduce a  battery storage duration $\tau_s$, after which we apply a Gaussian spin-selective $\pi/2$ pulse on L followed by a PFG (G$_{1z}$).
This ensures that there is no residual polarization of the load (L) spin.
We now introduce the discharging scheme $U_{XY}(\tau')$ between B and L.  Note that, the $U_{XY}$ propagator can be exactly implemented in the case of two-spin interaction. Finally, we measure the polarizations of both B and L spins after destroying the spurious coherences using a second PFG G$_\mathrm{z2}$, and thereby estimate their energies ${\tt e}_B$ and ${\tt e}_L$ respectively.  The experimental results of ${\tt e}_B$ and ${\tt e}_L$ are plotted versus $J_\mathrm{BL} \tau'$ in Fig. \ref{fig:CD1} (b).  
In our experiment, the load spin is beginning from a maximally mixed state instead of the ground state.  Therefore,  ${\tt e}_L$ starts with a value around 0.5 before raising towards the maximum value of 1.0 for $J_\mathrm{BL} \tau' =0.5$.  At this value of $J_\mathrm{BL} \tau'$, we vary the battery storage time $\tau_s$ and monitor the load energy ${\tt e}_L$.  The results are shown in Fig. \ref{fig:CD1} (c).  As expected, the data fits to an exponential decay function $e^{-\tau_s/T_s}$ (dashed line in Fig. \ref{fig:CD1} (c)) with an estimated battery storage time-constant $T_s \approx 200$ s.  This completes the demonstration of QCBL circuit.

\section{Summary and outlook} \label{Conclusion}
Considering the potential applications of quantum technologies, it is of great interest to study energy storage and usage at the quantum level.  In this context, there is a significant contemporary interest in studying quantum battery.
We investigated various aspects of  quantum battery using nuclear spin systems in star-topology molecules in the context of NMR architecture.  We first theoretically compared the  efficiency of the collective charging scheme (involving quantum correlation) with parallel (classical) scheme.  

Using NMR methods, we experimentally studied collective charging scheme in a variety of spin-systems, each having a single battery spin and a set of charger spins whose number $N$ ranged between 3 and 36.  By measuring the polarization of the battery spin, we estimated the battery energy and thereby established the  quantum advantage $\Gamma = \sqrt{N}$ of the collective charging scheme.

An important parameter to characterize a quantum battery is ergotropy, which quantifies the maximum amount of work that can be extracted from a quantum system via unitary methods. For each spin-system, we performed the experimental quantum state tomography and estimated the ergotropy of the battery spin and its evolution during charging. We observed the $\sqrt{N}$ quantum advantage in ergotropy as well.

By numerically evaluating  entanglement entropy and quantum discord for star-systems, we reconfirmed the established fact that the quantum advantage is realized via quantum correlation.
Therefore, we proposed using $\Gamma^2+1$ as an estimate for the size of the correlated cluster. In particular, for a 37 spin-system, we obtained an experimental value of $\Gamma \approx 6$, which in this case matched well with the expected number.  

We then addressed the issue of oscillatory charging wherein the battery starts discharging after overshooting the optimal charging duration.  To this end, we proposed a simple asymptotic charging method that involves iteratively re-energizing the charger with a suitable delay.  We experimentally demonstrated  asymptotic charging and determined the optimal delay range.  

Finally, we introduced a load spin to which the battery can deposit its energy after a suitable storage time, thus completing the complete charger-battery-load circuit.
Using a 38-spin system, we showed that the battery spin can store energy for up to two minutes and yet was able to transfer the stored energy to the load spin.

We believe this work paves the way for further methodology developments towards the practical aspects of quantum batteries.  Such developments may also contribute towards better understanding of quantum thermodynamics and its applications. One may also envisage an advanced circuit involving multiple elements such as quantum diodes, quantum transistors, and quantum heat engines, in addition to quantum batteries.  
\section*{Acknowledgements}
Authors 
acknowledges valuable discussions with Dr. Mir Alimuddin, Dr. Soham Pal, Dr. Rupak Bhattacharya, Krithika V. R., Priya Batra, Arijit Chatterjee, and Conan Alexander. 
TSM acknowledges funding from
DST/ICPS/QuST/2019/Q67.

\bibliographystyle{unsrtnat}
\bibliography{ref.bib}

\end{document}